\documentclass[reqno,12pt]{article}
\usepackage{amsmath,amsfonts,amssymb,amsthm,amstext,amscd,eucal,xcolor}
\usepackage{color}
\usepackage[all]{xy}
\usepackage{mathtools}
\newcommand{\mathsym}[1]{{}} 
\usepackage[colorlinks=true,
            linkcolor=blue,
            urlcolor=blue,
            citecolor=blue]{hyperref}
\usepackage{enumerate}

\usepackage{graphicx}
\usepackage{bm}
\usepackage{cite,hyperref}
\usepackage{mathtools}
\usepackage{graphicx}
\usepackage[active]{srcltx}
\makeatletter \@addtoreset{equation}{section}

\makeatletter\renewcommand\section{\@startsection {section}{1}{\z@}%
                                   {-3.5ex \@plus -1ex \@minus -.2ex}
                                   {2.3ex \@plus.2ex}%
                                   {\normalfont\large\bfseries}}
\renewcommand\subsection{\@startsection{subsection}{2}{\z@}%
                                     {-3.25ex\@plus -1ex \@minus -.2ex}%
                                     {1.5ex \@plus .2ex}%
                                     {\normalfont\bfseries}}

\DeclareMathAlphabet{\mathcal}{OMS}{cmsy}{b}{n}

\makeatother

\newcommand{\email}[1]{\footnote{E-mail: \href{mailto:#1}{#1}}}

\begin{document}

\title{\bf\Large{ Adler-Bell-Jackiw anomaly in VSR electrodynamics }}

\author{ R.~Bufalo\email{rodrigo.bufalo@ufla.br} $^{1}$, M.~Ghasemkhani\email{ghasemkhani@ipm.ir} $^{2}$, and A.~Soto\email{arsoto1@uc.cl} $^{3}$ \vspace{0.3cm}\\
\textit{$^{1}$ \small Departamento de F\'isica, Universidade Federal de Lavras,}\\
\textit{ \small Caixa Postal 3037, 37200-900 Lavras, MG, Brazil}\\
\textit{\small $^{2}$ Department of Physics, Shahid Beheshti University, 1983969411, Tehran, Iran }\\
\textit{\small $^{3}$  Instituto de F\'isica, Pontificia Universidad de Cat\'olica de Chile,}\\
\textit{\small Av. Vicu\~na Mackenna 4860, Santiago, Chile}\\
}

\maketitle
\date{}

\begin{abstract}
In this paper, we examine the problem of anomalies of the fermionic currents in the context of the very special relativity (VSR).
We consider the VSR contributions to the triangle amplitude  $
\left\langle J_{5}^{\lambda} J^{\mu} J^{\nu} \right\rangle $, which allows the evaluation of the vector and axial Ward identities.
Actually, we observe that the VSR nonlocal effects respect the vector Ward identity, and it also contributes in a very interesting and unique way for the Adler-Bell-Jackiw anomaly.
\end{abstract}

\section{Introduction}

In the development of models in quantum field theories, one necessarily enforces a symmetry principle when establishing the physical degrees of freedom (fields) present in the physical system \cite{sundermeyer}.
A desired symmetry would leave the classical action invariant under a given transformation of the fields.
As we know, the local gauge symmetries have been the keystone for the establishment of the standard model of the particle physics, as well as most of the phenomenological models.
On the other hand, there are symmetries related to the physical content of the fields and their couplings, depending on the given model, and have a crucial role in their definition.
These two types of symmetries, when preserved at the classical (tree) level, have an interesting feature in the quantum realm, because they are not necessarily preserved in the transition to the quantum theory, which are known as anomalous symmetries \cite{bertlmann,Harvey:2005it,Bilal:2008qx}.

The study of anomalies in the gauge theories is extremely important in demonstrating its consistency and physical properties, such as unitarity and renormalizability, and they are expected to cancel. For instance, these cancellations are of central importance in constraining the fermionic content of the standard model.
In order to verify whether a symmetry is preserved at the quantum level, one must analyze the respective Ward identity (or Slavnov-Taylor identity in the non-Abelian case).

Remarkably, the full structure of the anomaly can be calculated exactly at the one-loop order, and in the case of the chiral anomaly, it is given by the famous Adler-Bell-
Jackiw (ABJ) anomaly, justifying the $\pi_0 \to \gamma \gamma$ decay \cite{Bell:1969ts,Adler:1969gk,Jackiw:1986dr,bertlmann}.
As an another physical application of the anomalies, we can refer to the detected signatures of the chiral anomaly in the magneto-transport measurements in Weyl semimetals \cite{Zhang:2016ufu}.

Naturally, the models describing the physics beyond the standard model should fulfill such strong consistency requirements as well.
In this paper, we shall consider the problem of the anomalies in a particular class of the models presenting
Lorentz symmetry violation \cite{ref53,Bluhm:2005uj,Ellis:2005wr,AmelinoCamelia:2008qg}.
Contrary to the common lore, Lorentz violating effects are not necessarily related to Planck scale physics, it is also possible to formulate such class of models from a phenomenological group theory point of view.

A Lorentz violating framework, preserving all the basic elements of special relativity, is the Cohen and Glashow very special relativity (VSR) \cite{Cohen:2006ky,Cohen:2006ir}.
The main aspect of the VSR proposal is that the laws of physics
are invariant under the (kinematical) subgroups of the Poincar\'e group.
In $(3+1)$--dimensional spacetime, there are two VSR subgroups, SIM(2) and HOM(2), that preserve the direction of a lightlike four-vector $n_{\mu}$ by scaling, transforming as $n \to e^{\varphi} n$ under boost in the z direction.
This feature implies that there is a preferred direction in the Minkowski spacetime, where Lorentz violating terms can be constructed as ratios of contractions of the vector $n_{\mu}$ with other kinematical vectors, for instance $n^{\mu}/(n.p)$ \cite{Cohen:2006ky}.

The simplest example of VSR models is the free scalar massless field, whose action is given by
\begin{equation}
S = \int d^\omega x~\tilde{\partial}_{\mu} \phi \tilde{\partial}^{\mu} \phi,
\end{equation}
where the field equation reads $\widetilde{\Box}\phi=\left(\Box -m^2 \right)\phi  $, with the wiggle derivative operator defined by $\tilde{\partial}_{\mu}=\partial_{\mu}+\frac{1}{2}\frac{m^{2}}{n.\partial}n_{\mu}$.
We observe that the Lorentz violation appears in a nonlocal form and the parameter $m$ sets the scale for the VSR effects. \footnote{Since $m$ can be related to the neutrino and/or photon mass, for the VSR fermion and gauge sectors, respectively, the VSR parameter is very small, $m \ll 1$.}
The VSR approach has been extended to the gauge theories, where many interesting theoretical and phenomenological aspects of VSR effects have been extensively discussed \cite{Dunn:2006xk,Cheon:2009zx,Alfaro:2013uva,Alfaro:2015fha,Nayak:2016zed,Alfaro:2017umk,Alfaro:2019koq,Bufalo:2019kea}.

Hence, since VSR gauge theories possess very interesting features, we formulate in the present work a gauge invariant Lagrangian density in order to examine the ABJ anomaly in the VSR electrodynamics.
There were some works investigating the ABJ anomaly in the different Lorentz violating models \cite{Banerjee:2001un,Martin:2005gt,Sadooghi:2006sx,Arias:2007xt,Salvio:2008ta,Mariz:2015yua,AlMasri:2019uzr}, but not yet in the VSR context \footnote{After the completion of the present work a new manuscript presented an analysis of the Axial anomaly in VSR from a different point of view, obtaining different results \cite{Alfaro:2020zwo}.}.
Although, some aspects of the axial anomaly have been examined in the VSR-inspired Schwinger model \cite{Alfaro:2019snr}.

We start Sec.~\ref{sec2} by reviewing the main aspects of the VSR gauge invariance, and the related classical conserved currents in the VSR electrodynamics.
 The study of the axial anomaly in the VSR setup can be well motivated due to engendering nonlocal couplings as well as generating a massive mode in the propagator, which possibly will render nontrivial modifications.
Thus, it is valid to ask how the chiral anomaly behaves in the presence of such massive mode and nonlocal couplings coming from Lorentz violating effects.
In Sec.~\ref{sec3}, we consider the axial anomaly in terms of the triangle amplitude.
Since this amplitude needs to be regularized, we follow the 't Hooft-Veltman rule to perform algebraic manipulations with $\gamma_5$ within the dimensional regularization method.
We first evaluate the VSR contributions to the vector Ward identity.
Next, we discuss the VSR effects in the axial Ward identity, related to the ABJ anomaly.
In Sec.~\ref{conc}, we summarize the results, and present our final remarks.

\section{Anomalies in the VSR electrodynamics}
\label{sec2}

In order to discuss the ABJ anomaly in the context of VSR electrodynamics, we shall consider the following
Lagrangian density
\begin{equation} \label{eq1}
\mathcal{L}_{\textrm{vsr}}=\overline{\psi}\left(i\gamma^{\mu}\nabla_{\mu}-m_{e}\right)\psi,
\end{equation}
which establishes the dynamics of the fermions minimally coupled to an external electromagnetic potential $A_{\mu}$.
The VSR gauge coupling is defined by means of the covariant derivative as below
\begin{equation}
\nabla_{\mu}\psi=D_{\mu}\psi+\frac{1}{2}\frac{m^{2}n_{\mu}}{\left(n.D\right)}\psi,
\end{equation}
written in terms of the ordinary covariant derivative $D_{\mu}=\partial_{\mu}-ieA_{\mu}$, where $e$ is the gauge coupling,
and it reproduces the wiggle derivative $\tilde{\partial}_{\mu}=\partial_{\mu}+\frac{1}{2}\frac{m^{2}n_{\mu}}{\left(n.\partial\right)}$ in the noninteracting limit.

In our analysis of the problem of anomalies, in terms of the Ward identity, we shall first study the conserved currents in the classical level.
In this model, we have two conserved currents related to the gauge and chiral symmetries.
We can obtain these currents from the fermionic field equations
\begin{equation}
\left(i\gamma^{\mu}\nabla_{\mu}-m_{e}\right)\psi =0,
\end{equation}
and its conjugated. Thus, from straightforward manipulations, we obtain a vector and axial-vector currents, respectively \cite{Alfaro:2019snr}
\begin{align} \label{eq2}
J^{\mu} & =\overline{\psi}\gamma^{\mu}\psi+\frac{m^{2}}{2}\left(\frac{1}{n.D^{\dagger}}\overline{\psi}\right)\displaystyle{\not}nn^{\mu}\left(\frac{1}{n.D}\psi\right)  ,\\
J_{5}^{\mu} & = \overline{\psi}\gamma^{\mu}\gamma_{5}\psi+\frac{m^{2}}{2}\left(\frac{1}{n.D^{\dagger}}\overline{\psi}\right)\displaystyle{\not}nn^{\mu}\gamma_{5}\left(\frac{1}{n.D}\psi\right) .\label{eq3}
\end{align}
These two gauge invariant currents are classically conserved in the usual sense, i.e. $\partial_\mu J^{\mu} =0$ and $\partial_\mu J_{5}^{\mu} =2im_e J_5$, with $J_{5}  =  \overline{\psi} \gamma_{5}\psi$, in the chiral limit $m_{e}\rightarrow 0$.
The vector current \eqref{eq2} is invariant under the local transformations $\psi\rightarrow e^{ie\chi(x)}\psi,~\bar\psi\rightarrow e^{-ie\chi(x)}\bar\psi,~A_{\mu}\rightarrow A_{\mu}-\partial_{\mu}\chi(x)$, while
the axial-vector current \eqref{eq3} is invariant under the global chiral transformation $\psi\rightarrow e^{i \alpha\gamma^{5} }\psi,~\bar\psi\rightarrow \bar\psi e^{i \alpha\gamma^{5}}$.

An interesting point is that, since VSR effects can generate a massive mode for the fermionic field, how the chiral symmetry stands in this case, is it preserved or violated?
For that matter, we shall verify the validity of the chiral current conservation law, by means of the Ward identity, in the VSR framework.

In summary, we will study the ABJ anomaly of these currents in terms of the triangle amplitude, where the main object of analysis is the 3-point function
\footnote{It is important to emphasize that the full structure of the anomaly is given by the triangle graph \cite{Adler:1969er}.
This proof for QED uses arguments based on the renormalization program, showing  that higher-order  graphs contribute solely to the renormalization of fields and parameters.
Hence, since VSR electrodynamics is renormalizable  \cite{Alfaro:2015fha}, arguments analogous to the the proof can be carried through, establishing that the triangle graph also provides the full structure of the anomaly for VSR.}
\begin{equation}\label{eq4}
\left\langle J_{5}^{\lambda}\left(q\right)J^{\mu}\left(k_{1}\right)J^{\nu}\left(k_{2}\right)\right\rangle.
\end{equation}
This amplitude will be used in the evaluation of  the vector Ward identity $k_{1 \mu} \left\langle J_{5}^{\lambda}\left(q\right)J^{\mu}\left(k_{1}\right)J^{\nu}\left(k_{2}\right)\right\rangle  =?$, as well as the axial-vector Ward identity $q_{\lambda}\left\langle J_{5}^{\lambda}\left(q\right)J^{\mu}\left(k_{1}\right)J^{\nu}\left(k_{2}\right)\right\rangle  =?$.
Besides engendering massive modes for the dynamical fields, VSR effects are also present in the minimal coupling between the gauge field and the fermionic fields in terms of the Lagrangian density \eqref{eq1} (see eqs.\eqref{eq4.1} and \eqref{eq4.2}).
For this reason, the currents \eqref{eq2} and \eqref{eq3} are modified by the VSR nonlocal terms, and will be used to construct the triangle amplitude \eqref{eq4}.


Since the main object of analysis in the ABJ anomaly is the 3-point function $\left\langle J_{5}^{\lambda} J^{\mu} J^{\nu} \right\rangle$, one should pay attention in how to incorporate the VSR nonlocal couplings into the amplitude.
We observe first that the currents \eqref{eq2} and \eqref{eq3}, unlike the usual QED, have a dependence in the gauge field $A$; thus, we see that VSR generates an infinite perturbative series of nonlocal vertices \cite{Dunn:2006xk}.
However, these VSR vertex functions are higher-order in the coupling constant $e$,  and therefore the insertion of these vertices into each current of the function $\left\langle J_{5}^{\lambda} J^{\mu} J^{\nu} \right\rangle$ contributes to loop corrections of this amplitude (i.e. only to internal lines).
Hence, the only VSR coupling effect to the triangle graph comes from leading terms in the coupling constant $e$ of the currents \eqref{eq2} and \eqref{eq3} (see eqs.\eqref{eq4.1} and \eqref{eq4.2}, respectively).

Hence, the necessary Feynman rules for our analysis can be readily determined from \eqref{eq1}, and the currents \eqref{eq2} and \eqref{eq3}, where the fermionic propagator reads
\begin{equation}
S\left(p\right)=\frac{i\left(\displaystyle{\not}\tilde{p}+m_{e}\right)}{p^{2}-\mu^{2}},
\end{equation}
where $\mu^2=m^2+m_e^2$ is the fermionic mass, and the vertex Feynman rules are presented as below
\begin{itemize}
\item The vector current vertex
\begin{align}\label{eq4.1}
\Lambda^{\mu} \left( p,p' \right)& = -i \gamma^{\mu} -i\frac{m^{2}}{2}\frac{\displaystyle{\not}n~n^{\mu}}{\left(n.p\right)\left(n.p'\right)} \equiv  -i\gamma^\alpha\mathcal{R}^{\mu }_{\alpha} \left( p,p' \right);
\end{align}

\item The axial-vector current vertex
\begin{equation}\label{eq4.2}
\Phi_{5}^{\mu} \left( p,p' \right)=-i\gamma^{\mu}\gamma_{5} -i\frac{m^{2}}{2}\frac{\displaystyle{\not}n~n^{\mu}\gamma_{5}}{\left(n.p\right)\left(n.p'\right)} \equiv
-i\gamma^\alpha\gamma_5 \mathcal{R}^{\mu }_{\alpha} \left( p,p' \right);
\end{equation}

\end{itemize}
We have also introduced an useful notation in Eqs.~\eqref{eq4.1} and \eqref{eq4.2} to encode the VSR nonlocal factor
\begin{equation}\label{eq4.4}
\mathcal{R}^{\mu }_{\alpha} \left( p,p' \right)=  \delta_{\alpha}^{\mu}+\frac{m^{2}}{2}\frac{n_{\alpha}n^{\mu}}{\left(n.p\right)\left(n.p'\right)}.
\end{equation}

In the next section, we shall proceed to the analysis and evaluation of a series of Ward identities. All of these are related to the triangle graph $
\left\langle J_{5}^{\lambda}\left(q\right)J^{\mu}\left(k_{1}\right)J^{\nu}\left(k_{2}\right)\right\rangle $.
In the evaluation of this amplitude, we shall use the dimensional regularization in order to regularize it.
However, since the amplitude involves a $\gamma_5$ matrix, it is necessary to use the 't Hooft-Veltman rule in order to correctly define the $\gamma_5$ within the dimensional regularization  \cite{tHooft:1972tcz,Novotny:1994yx}.



\begin{figure}[t]
\vspace{-0.3cm}
\includegraphics[height=8.5\baselineskip]{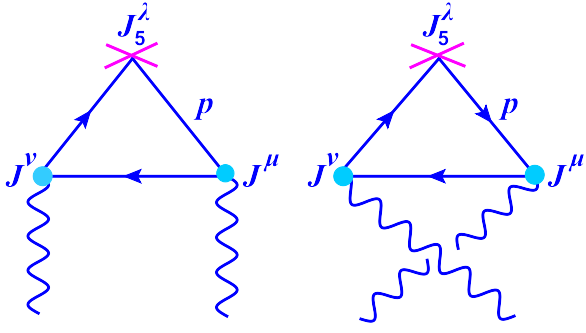}
 \centering\caption{ Relevant diagrams contributing to the ABJ anomaly. Cross symbol represents the axial-vector current,  whereas the circle symbol represents the vector current.}
\label{triangle_graphs}
\end{figure}

\section{Ward identities for the axial-vector vertex}
\label{sec3}

We proceed now to the discussion and evaluation of the amplitude related to the axial-vector vertex, which yields the so-called ABJ anomaly.
In this case, we shall consider two Ward identities, the vector and the axial one, related to the classical currents \eqref{eq2} and \eqref{eq3}, respectively.
At one-loop order, we have two diagrams contributing to the amplitude $ \left\langle J_{5}^{\lambda}\left(q\right)J^{\mu}\left(k_{1}\right)J^{\nu}\left(k_{2}\right)\right\rangle $, where the crossing symmetry is considered.
These contributions are depicted in Fig.~\ref{triangle_graphs}, in which the outgoing momenta of the photons are labeled by $k_{1}$ and $k_{2}$.
Hence, the Feynman expression of the relevant total amplitude is given by
\begin{align} \label{eq:15a}
\mathcal{A}_{5}^{\mu\nu\lambda}(q,k_{1},k_{2})=&-i \int_p\textrm{Tr}\biggl[S\left(p\right)\Phi_{5}^{\lambda} \left(p-q,p\right)S\left(p-q\right) \Lambda^{\nu}\left(p-k_{1},p-q\right) \cr
&\times S\left(p-k_{1}\right)\Lambda^{\mu}\left(p,p-k_{1}\right)\biggr]
+\Big(\begin{array}{c}
\mu\leftrightarrow\nu\\
k_{1}\leftrightarrow k_{2}
\end{array}\Big),
\end{align}
where the transferred momentum is $q=k_1+k_2$ and we have introduced the notation $\int_p = \int\frac{d^{\omega}p}{\left(2\pi\right)^{\omega}}$.
Moreover, in order to discuss the desired Ward identities, we can express the amplitude  \eqref{eq:15a} in its explicit form
\begin{align} \label{eq:15}
\mathcal{A}_{5}^{\mu\nu\lambda} (q,k_{1},k_{2})=&-i  \int_p\textrm{Tr}\frac{\left[\left(\displaystyle{\not}\tilde{p}+m_{e}\right)\gamma^{\sigma}\gamma_{5}\left(\widetilde{\left(\displaystyle{\not}p-\displaystyle{\not}q\right)}+m_{e}\right)\gamma^{\rho}\left(\widetilde{\left(\displaystyle{\not}p-\displaystyle{\not}k_{1}\right)}+m_{e}\right)\gamma^{\theta} \right]}{\left(p^{2}-\mu^{2}\right)\left(\left(p-q\right)^{2}-\mu^{2}\right)\left(\left(p-k_{1}\right)^{2}-\mu^{2}\right)}\cr
&\times\mathcal{R}_{\sigma}^{\lambda} \left(p-q,p\right)\mathcal{R}_{\rho}^{\nu}\left(p-k_{1},p-q\right)\mathcal{R}_{\theta}^{\mu}\left(p,p-k_{1}\right)+\left(\begin{array}{c}
\mu\leftrightarrow\nu\\
k_{1}\leftrightarrow k_{2}
\end{array}\right).
\end{align}

Some technical aspects that are worth recalling before the analysis:  the first step in computing the amplitude \eqref{eq:15} is the evaluation of the trace part.
For this, we should remember that the axial trace with 2 or an odd number of $\gamma$ matrices  vanishes, and also that $\textrm{Tr}\left[\gamma_{5}\gamma^{\rho}\gamma^{\alpha}\gamma^{\beta}\gamma^{\sigma}\right] =4i \epsilon^{\rho\alpha\beta\sigma}$, which follows from the identity
\begin{equation}
\gamma^{\beta}\gamma^{\theta}\gamma^{\lambda}=\eta^{\beta\theta}\gamma^{\lambda}-\eta^{\beta\lambda}\gamma^{\theta}
+\eta^{\theta\lambda}\gamma^{\beta}-i\epsilon^{\beta\theta\lambda\rho}\gamma_{\rho}\gamma_{5},
\end{equation}
Another important and useful identity is
\begin{equation}
\textrm{Tr}\Big(\gamma^{\mu_{1}}\gamma^{\mu_{2}}\ldots\gamma^{\mu_{n-1}}\gamma^{\mu_{n}}\Big)=
(-1)^{n}\textrm{Tr}\Big(\gamma^{\mu_{n}}\gamma^{\mu_{n-1}}\ldots\gamma^{\mu_{2}}\gamma^{\mu_{1}}\Big),
\end{equation}
that follows from the charge conjugation invariance, $C^{-1}\gamma^\mu C = - (\gamma^\mu)^T$ and $C^{-1}\gamma_5 C =  (\gamma_5)^T$, and it is valid for any number of $\gamma$ matrices.

In the discussion of the vector and axial Ward identities, the basic element is the amplitude  $\mathcal{A}_{5}^{\mu\nu\lambda}$, Eq. \eqref{eq:15}, that must be regularized.
However, since it is necessary to manipulate the $\gamma_{5}$ matrix in the trace computations, some issues are present when the dimensional regularization method is used \cite{Novotny:1994yx}.
Hence, in order to handle the problem of the definition of $\gamma_{5}$ in higher dimensions, we shall follow the 't Hooft-Veltman rule \cite{tHooft:1972tcz,Novotny:1994yx}.
This method consists in the splitting of the $\omega$ dimensional spacetime into two parts: a 4-dimensional (physical) and a $(\omega-4)$-dimensional subspace
\begin{equation} \label{eq:31a}
\int d^{\omega}p\to\int d^{\omega}P=\int d^{4}p\int d^{\omega-4}L.
\end{equation}
In this case, the internal momentum is expressed as below
\begin{equation} \label{eq:31}
\displaystyle{\not}P=\displaystyle{\not}p+\displaystyle{\not}L=\left(\gamma^{0}p_{0}+...+\gamma^{3}p_{3}\right)
+\left(\gamma^{4}L_{4}+...+\gamma^{\omega-1}L_{\omega-1}\right),
\end{equation}
where we have denoted the internal momentum as $L$ for the remaining  $\omega-4$ components.

Within the 't Hooft-Veltman rule, the $\gamma_{5}$ algebra is written as
\begin{align}
\left\{ \gamma_{5},\gamma^{\mu}\right\}  & =0,\quad\mu=0,1,2,3\\
\left[\gamma_{5},\gamma^{\mu}\right] & =0,\quad\mu=4,...,\omega-1,
\end{align}
and all other familiar rules are still valid, including the algebra
\begin{equation}
\left\{ \gamma^{\mu},\gamma^{\nu}\right\} =2g^{\mu\nu},\quad\mu,\nu=0,1,...,\omega-1,
\end{equation}
with the metric tensor components $g_{\mu\nu}=\textrm{diag}\left(+1,-1,...,-1\right)$.

We observe that all the external momenta and VSR null-vector, $\left(q,k_{1},k_{2},n\right)$, remain 4-dimensional.
This implies that the nonlocal factors are not affected by this variable separation, i.e. $\displaystyle{\not}\widetilde{P}=\widetilde{\left(\displaystyle{\not}p+\displaystyle{\not}L\right)}
=\displaystyle{\not}\widetilde{p}+\displaystyle{\not}L$ since $(n.P) = (n.p)$.
Moreover, we see that $\displaystyle{\not}L^2 =-L^2$ and $\displaystyle{\not}L \displaystyle{\not}p+ \displaystyle{\not}p\displaystyle{\not}L=0$.

Hence, with this set of results of the 't Hooft-Veltman rule, as well as the VSR invariant momentum integration (presented in the Appendix \ref{apA}), we are able to evaluate the vector Ward identity $k_{1\mu}\mathcal{A}_{5}^{\mu\nu\lambda}$ and the axial Ward identity $q_{\lambda}\mathcal{A}_{5}^{\mu\nu\lambda}$ in the VSR framework.


\subsection{Vector Ward Identity}

Let us evaluate the vector Ward identity $k_{1\mu}\mathcal{A}_{5}^{\mu\nu\lambda}$, which consists in the contraction of $k_{1\mu}$ into the expression \eqref{eq:15}.
Under this consideration, and using the dimensional regularization within the 't Hooft-Veltman prescription, we obtain
\begin{align} \label{eq:16}
k_{1\mu}\mathcal{A}_{5}^{\mu\nu\lambda}&=-i   \int_P \textrm{Tr}\frac{\left[\left(\displaystyle{\not}\tilde{P}+m_{e}\right)\gamma^{\sigma}\gamma_{5}\left( \widetilde{\left(\displaystyle{\not}P-\displaystyle{\not}q\right)}+m_{e}\right)\gamma^{\rho} \left( \widetilde{\left(\displaystyle{\not}P-\displaystyle{\not}k_1\right)}+m_{e}\right)\gamma^{\theta}\right]}{\left(P^{2}-\mu^{2}\right)\left(\left(P-q\right)^{2}-\mu^{2}\right)\left(\left(P-k_{1}\right)^{2}-\mu^{2}\right)}\cr
&\times\mathcal{R}_{\sigma}^{\lambda} \left(P-q,P\right)\mathcal{R}_{\rho}^{\nu}\left(P-k_{1},P-q\right)\left(k_{1\mu}\mathcal{R}_{\theta}^{\mu}\right)\left(P,P-k_{1}\right)+\left(\begin{array}{c}
\mu\leftrightarrow\nu\\
k_{1}\leftrightarrow k_{2}
\end{array}\right)
\end{align}
It is important to emphasize that we have indicated explicitly the splitting of spacetime dimensions, $\int d^{\omega}p\to\int d^{\omega}P=\int d^{4}p\int d^{\omega-4}L$, Eq.~\eqref{eq:31a}.
The complicated trace parts present in Eq.~\eqref{eq:16} can be casted into simpler forms, by using the following identities
\begin{align}
\displaystyle{\not}k_{1}\Big(\frac{1}{\displaystyle{\not}\widetilde{P}-m_e}\Big)=&1-\widetilde{\left(\displaystyle{\not}P-\displaystyle{\not}k_{1}\right)}\Big(\frac{1}{\displaystyle{\not}\widetilde{P} -m_e}\Big)
 +\frac{m_e}{\displaystyle{\not}\widetilde{P}-m_e}\cr
 &-\frac{m^{2}}{2}\displaystyle{\not}n\Big(\frac{1}{\displaystyle{\not}\widetilde{P}-m_e}\Big)\left(\frac{1}{n.\left(P-k_{1}\right)}-\frac{1}{\left(n.P\right)}\right),
\end{align}
and also
\begin{align}
\frac{1}{\widetilde{\left(\displaystyle{\not}P-\displaystyle{\not}q\right)}-m_e}\displaystyle{\not}k_{1}=&\frac{1}{\widetilde{\left(\displaystyle{\not}P-\displaystyle{\not}q\right)}-m_e}\widetilde{\left(\displaystyle{\not}P-\displaystyle{\not}k_{2}\right)}-1 -\frac{m_e}{\widetilde{\left(\displaystyle{\not}P-\displaystyle{\not}q\right)}-m_e} \cr
&+\frac{m^{2}}{2}\frac{1}{\widetilde{\left(\displaystyle{\not}P-\displaystyle{\not}q\right)}-m_e}
\displaystyle{\not}n\left(\frac{1}{n.\left(P-k_{2}\right)}-\frac{1}{n.\left(P-q\right)}\right).
\end{align}

We observe that the last term in the above identities, $m^2$ dependent (with nonlocal factors $1/(n.p_i)$) terms, are additional terms due to VSR in comparison to the usual algebra. This follows because of the result $\left(\tilde{P}+\tilde{q}\right) \neq \widetilde{\left(P+q\right)}$, remember that $\tilde{P}_\mu =P_\mu -\frac{m^2}{2} \frac{n_\mu}{(n.P)}$.

Finally, gathering all these results and substituting into \eqref{eq:16}, and making some algebraic manipulations within the 't Hooft-Veltman prescription, we find the simplified expression
\begin{align} \label{eq:20a}
k_{1\mu}\mathcal{A}_{5}^{\mu\nu\lambda}&=-i \frac{m^{2}}{2} \int_P\frac{\textrm{Tr}\left[\left({\displaystyle {\not}P-{\displaystyle {\not}k_{1}}}\right)\left({\displaystyle {\not}n}\right)\gamma_{5}\left({\displaystyle {\not}P-{\displaystyle {\not}q}}\right)\gamma^{\nu}\right]}{\left(\left(P-q\right)^{2}-m^{2}\right)\left(\left(P-k_{1}\right)^{2}-m^{2}\right)}\frac{n^{\lambda}}{\left(n.p\right)\left(n.\left(p-q\right)\right)}\cr
&+i \frac{m^{2}}{2} \int_P\frac{\textrm{Tr}\left[\left({\displaystyle {\not}P}\right)\left({\displaystyle {\not}n}\right)\gamma_{5}\left({\displaystyle {\not}P-{\displaystyle {\not}k_{2}}}\right)\gamma^{\nu}\right]}{\left(P^{2}-m^{2}\right)\left(\left(P-k_{2}\right)^{2}-m^{2}\right)}\frac{n^{\lambda}}{\left(n.p\right)\left(n.\left(p-q\right)\right)}\cr
&+i \frac{m^{2}}{2} \int_P \frac{\textrm{Tr}\left[\left({\displaystyle {\not}P+{\displaystyle {\not}k_{1}}}\right)\gamma^{\lambda}\gamma_{5}\left({\displaystyle {\not}P-{\displaystyle {\not}k_{2}}}\right)\left({\displaystyle {\not}n}\right)\right]}{\left(\left(P+k_{1}\right)^{2}-m^{2}\right)\left(\left(P-k_{2}\right)^{2}-m^{2}\right)}\frac{n^{\nu}}{\left(n.p\right)\left(n.\left(p-k_{2}\right)\right)}\cr
&-i \frac{m^{2}}{2} \int_P\frac{\textrm{Tr}\left[\left({\displaystyle {\not}P}\right)\gamma^{\lambda}\gamma_{5}\left({\displaystyle {\not}P-{\displaystyle {\not}q}}\right)\left({\displaystyle {\not}n}\right)\right]}{\left(P^{2}-m^{2}\right)\left(\left(P-q\right)^{2}-m^{2}\right)}\frac{n^{\nu}}{\left(n.p\right)\left(n.\left(p-k_{2}\right)\right)}
\end{align}
where we have considered the massless limit $m_{e}=0$ as a simplification.
It is interesting to observe that the cancelled terms in our development from eq.\eqref{eq:16} to eq.\eqref{eq:20a}, are due to the shift $p\to p+k_1$ (allowed in dimensional regularization); even the usual QED terms are canceled due to this change of variables (see that eq.\eqref{eq:20a} is purely a VSR effect, since it consists of $m^2$ terms).
However, as one can see, some terms are not cancelled under  shifts (due to the nonlocal factors).
Hence, in order to verify the vector WI, it is necessary to evaluate explicitly the momentum integration in eq.\eqref{eq:20a}.

The remaining terms in eq.\eqref{eq:20a} can now be readily evaluated by computing the trace parts. To simplify the momentum integrals, it is convenient to use the following property
 \begin{equation}
 \frac{1}{ n.\left(p+k_i\right)n.\left(p+k_j\right)} = \frac{1}{ n.\left(k_i-k_j\right)} \left(  \frac{1}{ n.\left(p+k_j\right)} - \frac{1}{ n.\left(p+k_i\right)}\right).
  \label{eq:11}
 \end{equation}
Hence, taking all of these steps into consideration, and solving the momentum integrals with the results provided in the Appendix \ref{apA}, we find from eq.\eqref{eq:20a} that the vector Ward identity reads
\begin{align} \label{eq:40}
k_{1\mu}\mathcal{A}_{5}^{\mu\nu\lambda}=0.
\end{align}
We see that the vector Ward identity \eqref{eq:40} is satisfied for the VSR axial-vector vertex.
This is already an interesting result, because it shows that the gauge structure of the VSR electrodynamics is well established at quantum level and hence the charge is conserved.
We proceed next to the evaluation of the axial Ward identity, and shall compute the VSR contributions to the ABJ anomaly.

\subsection{Axial Ward identity}

The axial Ward identity comes from the contraction  of \eqref{eq:15} with $q_\lambda$ in the axial-vector vertex, so that within the 't Hooft-Veltman prescription it can be cast as
\begin{align} \label{eq:25}
q_\lambda \mathcal{A}_{5}^{\mu\nu\lambda}=&-i   \int_P \textrm{Tr}\frac{\left[\left(\displaystyle{\not}\tilde{P}+m_{e}\right)\gamma^{\sigma}\gamma_{5}\left(\widetilde{\left(\displaystyle{\not}P-\displaystyle{\not}q\right)}+m_{e}\right)\gamma^{\rho}\left(\widetilde{\left(\displaystyle{\not}P-\displaystyle{\not}k_{1}\right)}+m_{e}\right)\gamma^{\theta} \right]}{\left(P^{2}-\mu^{2}\right)\left(\left(P-q\right)^{2}-\mu^{2}\right)\left(\left(P-k_{1}\right)^{2}-\mu^{2}\right)}\cr
&\times \left[ q_\lambda \mathcal{R}_{\sigma}^{\lambda} \left(P-q,P\right)  \right] \mathcal{R}_{\rho}^{\nu}\left(P-k_{1},P-q\right)\mathcal{R}_{\theta}^{\mu}\left(P,P-k_{1}\right)+\left(\begin{array}{c}
\mu\leftrightarrow\nu\\
k_{1}\leftrightarrow k_{2}
\end{array}\right).
\end{align}

Furthermore, in order to simplify the trace parts of the expression \eqref{eq:25}, we shall consider the following identity
\begin{align}
\displaystyle{\not}q\gamma_{5}  =&\gamma_{5}\left( \widetilde{\left(\displaystyle{\not}P -\displaystyle{\not}q\right)} -m_e\right)+ \left( \displaystyle{\not}\tilde{P}-m_e\right)\gamma_{5}-2\displaystyle{\not}L\gamma_{5}  \cr
&+2m_e \gamma_5
 +\frac{m^{2}}{2}\left(\frac{1}{n.\left(P-q\right)}-\frac{1}{n.P} \right)\gamma_{5}\displaystyle{\not}n,
\end{align}
that takes into account the 't Hooft-Veltman rule for the  spacetime splitting and $\gamma_{5}$ algebra, and also some properties from VSR.
After some straightforward manipulation, we find that
\begin{align}
\frac{1}{\displaystyle{\not}\tilde{P} -m_e }\displaystyle{\not}(\tilde{q}\gamma_{5})\frac{1}{\widetilde{\left(\displaystyle{\not}P-\displaystyle{\not}q\right)}-m_e } & =\frac{1}{\displaystyle{\not}\tilde{P} -m_e }\gamma_{5}+\gamma_{5}\frac{1}{\widetilde{\left(\displaystyle{\not}P-\displaystyle{\not}q\right)}-m_e } -2\frac{1}{\displaystyle{\not}\tilde{P} -m_e }(\displaystyle{\not}L\gamma_{5})\frac{1}{\widetilde{\left(\displaystyle{\not}P-\displaystyle{\not}q\right)}-m_e }  \cr
  & +\frac{m^{2}}{2}\frac{1}{\displaystyle{\not}\tilde{P} -m_e }(\gamma_{5}\displaystyle{\not}n)\frac{1}{\widetilde{\left(\displaystyle{\not}P-\displaystyle{\not}q\right)} -m_e }\left(\frac{1}{n.\left(P-q\right)}-\frac{1}{n.P} \right) \cr
  & +2m_e \frac{1}{\displaystyle{\not}\tilde{P} -m_e }\gamma_{5}  \frac{1}{\widetilde{\left(\displaystyle{\not}P-\displaystyle{\not}q\right)} -m_e}.
\end{align}

Hence, by considering these results, we are able to rewrite the axial Ward identity \eqref{eq:25} in a simplified form \footnote{Some terms vanished since
it is impossible to form a two-index pseudotensor which depends on only one vector.}
\begin{align}
q_{\lambda}\mathcal{A}_{5}^{\mu\nu\lambda}=&-i  \int_P \frac{\textrm{Tr}
\left[ \left(\widetilde{ \displaystyle{\not}u  }+m_{e}\right)\gamma_{5}\gamma^{\beta} \left(\displaystyle{\not}\tilde{P}+m_{e}\right) \gamma^{\alpha}\right]}{\left(P^{2}-\mu^{2}\right)\left(u^{2}-\mu^{2}\right)} \mathcal{R}_{\alpha}^{\mu}\left(u,p\right)\mathcal{R}_{\beta}^{\nu}\left(p,w\right) \cr
&-i     \int_P \frac{\textrm{Tr}\left[\gamma_{5}\left(\widetilde{ \displaystyle{\not}w }+m_{e}\right)\gamma^{\beta}\left(\displaystyle{\not}\tilde{P}+m_{e}\right)\gamma^{\alpha}\right]}{\left(P^{2}-\mu^{2}\right)\left(w^{2}-\mu^{2}\right)} \mathcal{R}_{\alpha}^{\mu}\left(u,p\right)\mathcal{R}_{\beta}^{\nu}\left(p,w\right) \cr
&+2i  \int_P \frac{\textrm{Tr}\left[\left(\widetilde{ \displaystyle{\not}u }+m_{e}\right)\displaystyle{\not}L\gamma_{5}\left(\widetilde{ \displaystyle{\not}w }+m_{e}\right)\gamma^{\beta} \left(\displaystyle{\not}\tilde{P}+m_{e}\right)\gamma^{\alpha}\right]}{\left(P^{2}-\mu^{2}\right)\left(u^{2}-\mu^{2}\right)\left(w^{2}-\mu^{2}\right)} \mathcal{R}_{\alpha}^{\mu}\left(u,p\right)\mathcal{R}_{\beta}^{\nu}\left(p,w\right) \cr
&-2i m_{e} \int_P \frac{\textrm{Tr}\left[\left({\displaystyle \widetilde{ {\displaystyle {\not}u } }+m_{e}}\right)\gamma_{5}\left(\widetilde{{\displaystyle {\not}w }}  +m_{e}\right){\displaystyle \gamma^{\beta}\left(\widetilde{{\displaystyle {\not}P}}+m_{e}\right)\gamma^{\alpha}}\right]}{\left( u^{2}-\mu^{2}\right)\left(w^{2}-\mu^{2}\right)\left(P^{2}-\mu^{2}\right)} \mathcal{R}_{\alpha}^{\mu}\left(u,p\right)\mathcal{R}_{\beta}^{\nu}\left(p,w\right)\cr
&+\left(\begin{array}{c}
\mu\leftrightarrow\nu\\
k_{1}\leftrightarrow k_{2}
\end{array}\right).
\label{eq:30}
\end{align}
where we have defined $u=P+k_1$ and $w=P-k_2$, and the symmetrization symbol $\left(\begin{array}{c}
\mu\leftrightarrow\nu\\
k_{1}\leftrightarrow k_{2}
\end{array}\right)$ holds for all the terms in \eqref{eq:30}.
The evaluation of the trace parts and momentum integration of \eqref{eq:30} is a lengthy, tedious, but straightforward task.
To this end, one should apply first the spacetime decomposition as presented in the Eq.~\eqref{eq:31}, then use the identity $\textrm{Tr}\big(\gamma_{5}\gamma^{\xi}\gamma^{\alpha}\gamma^{\chi}\gamma^{\beta}\big)=4i\epsilon^{\xi\alpha\chi\beta}$ to compute the trace parts.

Furthermore, the integrals in the $(p,L)$ momentum variables are evaluated with aid of the results presented in the Appendix \ref{apA}.
Finally, we gather all the contributions to the complete expression of the axial Ward identity as
\footnote{ One can see that our result \eqref{eq:41} contains VSR contributions, coming from the presence of the nonlocal factors. This is in contrast to the outcomes of ref.\cite{Alfaro:2020zwo}, where no VSR effects are found in the axial Ward identity. It is surprising to observe this discrepancy, because the starting point in both analysis is the same eq.\eqref{eq:25}. Actually, the author considered explicitly the limit of ``large $q$'' in the path integral analysis of the anomaly \cite{Alfaro:2020zwo}. It is obvious that, under this limit of large momentum, all the VSR nonlocal factors $1/(n.q)$ vanish. Perhaps, similar approximations were considered throughout the perturbative analysis, justifying the absence of VSR effects in the axial anomaly found in \cite{Alfaro:2020zwo}.}
\begin{align} \label{eq:41}
q_{\lambda}\mathcal{A}_{5}^{\mu\nu\lambda}&=\frac{i }{2\pi^{2}}\varepsilon^{\lambda\sigma\mu\nu}k_{2\lambda}k_{1\sigma}+\frac{i m_{e}^{2}}{\pi^{2}}\epsilon^{\xi\nu\mu\rho}k_{2\xi}k_{1\rho}\int dxdy\frac{1}{M_{1}}\cr
&+\frac{i m_{e}^{2}m^{2}}{2\pi^{2}}\left(\epsilon^{\xi\nu\mu\rho}n_{\xi}k_{2\rho}-\epsilon^{\xi\nu\sigma\rho}n_{\sigma}k_{2\xi}k_{1\rho}n^{\mu}\frac{1}{\left(n.k_{1}\right)}\right)\int dxdy\frac{1}{\left(n\cdot r_1\right)}\left[\frac{1}{M_{1}}-\frac{1}{M_{1}+r_1^{2}}\right]\cr
&-\frac{i m_{e}^{2}m^{2}}{2\pi^{2}}\left(\epsilon^{\xi\nu\mu\rho}n_{\xi}k_{1\rho}+\epsilon^{\xi\mu\sigma\rho}n_{\sigma}k_{2\xi}k_{1\rho}n^{\nu}\frac{1}{\left(n.k_{2}\right)}\right)\int dxdy\frac{1}{\left(n.r_2\right)}\left[\frac{1}{M_{2}}-\frac{1}{M_{2}+r_2^{2}}\right]\cr
&-\frac{i m_{e}^{2}m^{2}}{2\pi^{2}}\left(\epsilon^{\xi\nu\mu\rho}n_{\xi}q_{\rho}+\frac{\epsilon^{\xi\mu\sigma\rho}n_{\sigma}k_{2\xi}k_{1\rho}n^{\nu}}{\left(n.k_{2}\right)}+\frac{\epsilon^{\xi\nu\sigma\rho}n_{\sigma}k_{2\xi}k_{1\rho}n^{\mu}}{\left(n.k_{1}\right)}\right) \int \frac{dxdy }{\left(n\cdot s\right)}\left[\frac{1}{M_{3}}-\frac{1}{M_{3}+s^{2}}\right] \cr
&+\frac{i m^{2}}{4\pi^{2}}\frac{\epsilon^{\xi\nu\chi\mu}n_{\chi}k_{2\xi}}{\left(n.k_{2}\right)}\int dx\left[\ln\left[\frac{\mu^{2}-xk_{2}^{2}}{\mu^{2}}\right]+\left(1-\frac{1}{x}\right)\ln\left[1+\frac{x^{2}k_{2}^{2}}{\mu^{2}-xk_{2}^{2}}\right]\right]\cr
&-\frac{i m^{2}}{4\pi^{2}}\frac{\epsilon^{\xi\nu\chi\mu}n_{\chi}k_{1\xi}}{\left(n\cdot k_{1}\right)}\int dx\left[\left(1-\frac{1}{x}\right)\ln\left[1+\frac{x^{2}k_{1}^{2}}{\mu^{2}-xk_{1}^{2}}\right]+\ln\left[\frac{\mu^{2}-xk_{1}^{2}}{\mu^{2}}\right]\right]\cr
&+\frac{i m^{2}}{4\pi^{2}}\frac{\epsilon^{\lambda\alpha\mu\nu}n_{\lambda}\left(k_{1}-k_{2}\right)_{\alpha}}{\left(n\cdot q\right)}\int dx\frac{1}{x}\ln\left[1+\frac{x^{2}q^{2}}{\mu^{2}-xq^{2}}\right]  ,
\end{align}
with
\begin{align}
r_{1,2}&=-yk_{1,2}-\left(1-x-y\right)q, \\
s&=-yk_{1}-\left(1-x-y\right)k_{2}\\
M_{1,2}&=\mu^{2}-yk_{1,2}^{2}-\left(1-x-y\right)q^{2},\\
 M_{3}&=\mu^{2}-yk_{1}^{2}-\left(1-x-y\right)k_{2}^{2}.
\end{align}
We notice that the first two terms of \eqref{eq:41} correspond to the results of QED with massive fermions \cite{Adler:1969gk}, $q_{\lambda}\mathcal{A}_{5}^{\mu\nu\lambda}=\frac{i }{2\pi^{2}}\varepsilon^{\lambda\sigma\mu\nu}k_{2\lambda}k_{1\sigma}+2m_e \mathcal{A}_{5}^{\mu\nu }$, where $\mathcal{A}_{5}^{\mu\nu }=\frac{i m_{e}}{2\pi^{2}}\epsilon^{\xi\nu\mu\rho}k_{2\xi}k_{1\rho}\int dxdy\frac{1}{M_{1}}$.
It is interesting to observe that the expression \eqref{eq:41} shows that the VSR contributes to the axial anomaly in a very distinct way.
The tensor structure is quite different from the ABJ term, presenting novel contractions of the external 4-momentum $(k_1,k_2)$ with the VSR null vector $n_\alpha$ (besides the presence of the nonlocal terms $1/(n.k_i)$).

\section{Final remarks}
\label{conc}

In this paper, we have analyzed the Adler-Bell-Jackiw anomaly for Dirac fermions in the context of the very special relativity.
The main goal of the proposed study was to examine the behavior of known conservation laws in the quantum realm of a Lorentz violating model, by analyzing the radiative corrections related to the VSR nonlocal gauge couplings.

We started by establishing and discussing the classical conserved currents, vector and axial-vector ones, that are modified by VSR contributions.
Since the anomaly is fully determined by the triangle graphs for renormalizable field theories, we have established the amplitude $\left\langle J_{5}^{\lambda}\left(q\right)J^{\mu}\left(k_{1}\right)J^{\nu}\left(k_{2}\right)\right\rangle$ in the VSR framework, and considered the respective Ward identities.
We have also discussed in detail how the VSR nonlocal couplings could be incorporated into this amplitude, and how the VSR modified vertex  $\left\langle A \bar{\psi}\psi\right\rangle$ is the only contribution at the leading order.

In the evaluation of the amplitude $\left\langle J_{5}^{\lambda} J^{\mu} J^{\nu} \right\rangle$, it is required the use of the 't Hooft-Veltman rules, since it involved the algebraic manipulation of the $\gamma_5$ in the context of dimensional regularization.
In the case of the vector Ward identity, we observe that the VSR effects do not change its structure, showing that it is satisfied, corroborating the fact that the gauge symmetry is well formulated in the VSR electrodynamics.
On the other hand, in the case of the axial Ward identity, we find that VSR modify the known QED Ward identity for arbitrary values of the parameter $m^2$.

Hence, while VSR effects respect the vector Ward identity, its effect upon the axial Ward identity is very different:   VSR contributes to the axial anomaly in an interesting way. The tensor structure is quite different from the ABJ term, presenting novel contractions of the external 4-momentum $(k_1,k_2)$ with the VSR null vector $n_\alpha$.

 Since there is a close relation among the ABJ anomaly with the $\pi_0 \to \gamma \gamma$ decay \cite{Adler:1969gk,Jackiw:1986dr}, it is a natural extension to examine the VSR effects upon this type of anomalous processes and find how the VSR  nonlocal effects contribute to its decay width.
This is currently under development and will be reported elsewhere.

 \subsection*{Acknowledgements}

R.B. acknowledges partial support from Conselho
Nacional de Desenvolvimento Cient\'ifico e Tecnol\'ogico (CNPq Projects No. 305427/2019-9 and No. 421886/2018-8) and Funda\c{c}\~ao de
Amparo \`a Pesquisa do Estado de Minas Gerais (FAPEMIG Project No. APQ-01142-17).

\appendix

\section{Useful integrals}
\label{apA}

In order to cope with the momentum integral in VSR, we use the Mandelstam-Leibbrant prescription extended to VSR
\cite{Alfaro:2017umk}
 \begin{align}
\int d^{\omega}q \frac{1}{(q^2 + 2 q.p - m^2)^a} \frac{1}{(n. q)^b}
 = (- 1)^{a + b} i \pi^{\frac{\omega}{2}} (- 2)^b \frac{\Gamma (a + b - \frac{\omega}{2})}{\Gamma
(a) \Gamma (b)} (\bar{n}.p)^b \int^1_0 d t~t^{b - 1} \frac{1}{ \Delta^{a + b - \frac{\omega}{2}}},
\label{eq:8}
\end{align}
where $\Delta = m^2 + p^2 - 2 (n.p) (\bar{n}.p) t$, and $\bar{n}$ is a new null vector ($\bar{n}^2 = 0 $) with the property $(n.\bar{n}) = 1 $, whose explicit form must be provided within the VSR framework \cite{Alfaro:2017umk}.
Taking into account the properties such as
reality, right scaling $(n,\bar{n}) \to (\lambda n, \lambda^{-1}\bar{n})$  and being dimensionless \cite{Alfaro:2017umk}, we find a VSR invariant form as $\bar{n}_\mu=\frac{p_\mu}{(n.p)}-\frac{p^2 n_\mu}{2 (n.p)^2}  $.
The remaining useful integrals with $q^\mu$ and $q^\mu q^\nu $ in the numerator can be obtained by direct derivation of \eqref{eq:8} in relation to $p_\mu$ and $ p_\mu p_\nu$, respectively.

\global\long\def\link#1#2{\href{http://eudml.org/#1}{#2}}
 \global\long\def\doi#1#2{\href{http://dx.doi.org/#1}{#2}}
 \global\long\def\arXiv#1#2{\href{http://arxiv.org/abs/#1}{arXiv:#1 [#2]}}
 \global\long\def\arXivOld#1{\href{http://arxiv.org/abs/#1}{arXiv:#1}}


\end{document}